\documentclass[10pt,conference]{IEEEtran}
\IEEEoverridecommandlockouts

\usepackage{cite}
\usepackage{amsmath,amssymb,amsfonts}
\usepackage{algorithmic}
\usepackage{graphicx}
\usepackage{textcomp}
\usepackage{tabularx}
\usepackage[table,xcdraw]{xcolor}
\usepackage{makecell}
\usepackage{array}

\usepackage{booktabs}
\def\BibTeX{{\rm B\kern-.05em{\sc i\kern-.025em b}\kern-.08em
    T\kern-.1667em\lower.7ex\hbox{E}\kern-.125emX}}
\begin{document}
\begin{sloppy}

\title{How Developers Interact with AI: A Taxonomy of Human-AI Collaboration in Software Engineering}

\author{\IEEEauthorblockN{Christoph Treude}
\IEEEauthorblockA{\textit{School of Computing and Information Systems} \\
\textit{Singapore Management University}\\
Singapore\\
ctreude@smu.edu.sg}
\and
\IEEEauthorblockN{Marco A.~Gerosa}
\IEEEauthorblockA{\textit{School of Informatics, Computing, and Cyber Systems} \\
\textit{Northern Arizona University}\\
Flagstaff, AZ, United States\\
marco.gerosa@nau.edu}}

\maketitle

\begin{abstract}
Artificial intelligence (AI), including large language models and generative AI, is emerging as a significant force in software development, offering developers powerful tools that span the entire development lifecycle. Although software engineering research has extensively studied AI tools in software development, the specific types of interactions between developers and these AI-powered tools have only recently begun to receive attention. Understanding and improving these interactions has the potential to enhance productivity, trust, and efficiency in AI-driven workflows. In this paper, we propose a taxonomy of interaction types between developers and AI tools, identifying eleven distinct interaction types, such as auto-complete code suggestions, command-driven actions, and conversational assistance. Building on this taxonomy, we outline a research agenda focused on optimizing AI interactions, improving developer control, and addressing trust and usability challenges in AI-assisted development. By establishing a structured foundation for studying developer-AI interactions, this paper aims to stimulate research on creating more effective, adaptive AI tools for software development.
\end{abstract}


\section{Introduction}

Artificial intelligence (AI) is rapidly transforming software development, bringing new capabilities and efficiency to every stage of the development lifecycle~\cite{russo2024navigating}. As AI tools become increasingly sophisticated, they offer developers a diverse range of support options, from autocompletion and code generation to documentation, testing, and project management~\cite{durrani2024decade}. Related work on developer interactions with traditional tools has established the importance of usability and integration~\cite{minelli2014visualizing}, but there is limited research specifically on AI-powered tools\cite{brown2024identifying}. The paper aims to address this gap. 

This emerging focus on AI-powered tools brings with it an opportunity to explore how the interaction of the tools can be tailored to better align with developers' needs, ultimately enhancing adoption, trust, and productivity~\cite{murillo2024understanding}. The diversity of interaction types -- from auto-complete code suggestions to conversational exchanges~\cite{ross2023programmer} -- introduces significant complexity in designing and evaluating these tools. Without a structured framework for analyzing these interactions, researchers and tool designers lack a common vocabulary and conceptual model for developer-AI interfaces. This fragmentation hinders efforts to optimize how developers engage with AI tools, identify best practices, understand usage patterns, and address common challenges across different tools and contexts. A comprehensive taxonomy of developer-AI interactions can help bridge this gap by providing a foundation for empirical studies, enabling systematic tool evaluation, and guiding the development of AI assistance features. Furthermore, as AI capabilities continue to evolve rapidly, such a taxonomy becomes essential for tracking how interaction patterns adapt and emerge over time, ensuring that new tools and features align with developers' actual needs and workflows.

In this paper, we propose a taxonomy of interaction types between human developers and AI tools. To the best of our knowledge, this is the first taxonomy to specifically examine human-AI interaction types within the context of software engineering, a domain characterized by the need to manage diverse types of artifacts (e.g., code, bug reports, pull requests), navigate complex human collaboration dynamics, and address rapidly evolving development workflows. Building on this taxonomy, we outline a research agenda that highlights areas for further exploration, including optimizing interaction styles, improving developer control, and addressing trust and usability in AI-driven development. This paper aims to catalyze further research and discussion on creating more effective and adaptive AI tools for software development.

\section{Interaction Types}

\begin{table*}[ht]
\centering
\caption{Characterization of Developer-AI Interaction Types}
\label{tab:interaction-types}
\begin{tabularx}{\textwidth}{>{\raggedright\arraybackslash}p{3.5cm}|>{\raggedright\arraybackslash}p{6cm}>{\raggedright\arraybackslash}p{3.5cm}>{\raggedright\arraybackslash}p{3.5cm}}
\toprule
\rowcolor[HTML]{EFEFEF} 
\textbf{Type} & \textbf{Trigger and AI Response} & \textbf{Developer Response and Output} & \textbf{Example} \\ 
\midrule
\rowcolor[HTML]{FFFFFF} 
Auto-Complete Code Suggestions & 
\textit{Trigger:} Automatic based on typing context. \newline 
\textit{AI Response:} Suggestions appear as ghost text or pop-ups. & 
\textit{Developer Response:} Accept, scroll, or dismiss. \newline 
\textit{Output:} Suggestions. & 
Typing "def calculate\_area" prompts a function body suggestion in GitHub Copilot~\cite{bird2022taking}. \\ 
\midrule
\rowcolor[HTML]{EFEFEF} 
Command-Driven Actions & 
\textit{Trigger:} Explicit command input (e.g., \texttt{copilot:summary}). \newline 
\textit{AI Response:} Generates specified output (e.g., summary or documentation). & 
\textit{Developer Response:} Review, edit, finalize. \newline 
\textit{Output:} Actions. & 
Using \texttt{copilot:summary} generates a pull request summary in GitHub~\cite{xiao2024generative}. \\ 
\midrule
\rowcolor[HTML]{FFFFFF} 
Conversational Assistance & 
\textit{Trigger:} Question or issue posed in a chat interface. \newline 
\textit{AI Response:} Step-by-step guidance, explanations, or snippets. & 
\textit{Developer Response:} Copy, adapt, or ask follow-ups. \newline 
\textit{Output:} Explanations. & 
ChatGPT explains sorting a list of dictionaries~\cite{xiao2024devgpt}. \\ 
\midrule
\rowcolor[HTML]{EFEFEF} 
Contextual Recommendations & 
\textit{Trigger:} Interprets contextual cues (e.g., file type). \newline 
\textit{AI Response:} Suggests libraries, patterns, or improvements. & 
\textit{Developer Response:} Evaluate, accept, modify. \newline 
\textit{Output:} Suggestions. & 
Sourcegraph Cody suggests database libraries for relevant files~\cite{hartman2024ai}. \\ 
\midrule
\rowcolor[HTML]{FFFFFF} 
Selection-Based Enhancements & 
\textit{Trigger:} Highlighting specific code segments. \newline 
\textit{AI Response:} Provides refactored code, explanations, or tests. & 
\textit{Developer Response:} Review, incorporate, modify. \newline 
\textit{Output:} Actions. & 
Sourcery refactors highlighted functions~\cite{dwivedi2024more}. \\ 
\midrule
\rowcolor[HTML]{EFEFEF} 
Explicit UI Actions & 
\textit{Trigger:} Button or icon clicked in the IDE. \newline 
\textit{AI Response:} Displays flagged issues, reports, or documentation. & 
\textit{Developer Response:} Review, refine, incorporate. \newline 
\textit{Output:} Actions. & 
Security scans in IDEs flag vulnerabilities for review~\cite{pudari2023copilot}. \\ 
\midrule
\rowcolor[HTML]{FFFFFF} 
Comment-Guided Prompts & 
\textit{Trigger:} Descriptive comments written by the developer. \newline 
\textit{AI Response:} Generates code beneath the comment. & 
\textit{Developer Response:} Review, adjust, verify. \newline 
\textit{Output:} Actions. & 
Writing "// Convert list of strings to uppercase" prompts code in GitHub Copilot~\cite{bird2022taking}. \\ 
\midrule
\rowcolor[HTML]{EFEFEF} 
Event-Based Triggers & 
\textit{Trigger:} Workflow events (e.g., commits or pull requests). \newline 
\textit{AI Response:} Reports issues or performs checks. & 
\textit{Developer Response:} Review and address identified issues. \newline 
\textit{Output:} Actions. & 
GitLab Auto DevOps scans pull requests for vulnerabilities~\cite{shi2024greening}. \\ 
\midrule
\rowcolor[HTML]{FFFFFF} 
Shortcut-Activated Commands & 
\textit{Trigger:} Shortcut keys pressed. \newline 
\textit{AI Response:} Provides suggestions or documentation in an overlay. & 
\textit{Developer Response:} Evaluate, integrate, or dismiss. \newline 
\textit{Output:} Suggestions. & 
Shortcuts in IntelliJ IDEA open Copilot suggestions~\cite{zhang2023demystifying}. \\ 
\midrule
\rowcolor[HTML]{EFEFEF} 
File-Aware Suggestions & 
\textit{Trigger:} Recognizes file type or directory context. \newline 
\textit{AI Response:} Suggests templates or configuration options. & 
\textit{Developer Response:} Review, accept, or adapt. \newline 
\textit{Output:} Suggestions. & 
CodeWhisperer suggests test templates for \texttt{.test} files~\cite{mihaljevic2024analysis}. \\ 
\midrule
\rowcolor[HTML]{FFFFFF} 
Automated API Responses & 
\textit{Trigger:} API calls or webhooks triggered by events. \newline 
\textit{AI Response:} Provides reports or release notes. & 
\textit{Developer Response:} Review, modify, and integrate. \newline 
\textit{Output:} Reports. & 
CodeClimate analyzes pull requests and reports quality issues~\cite{hu2019improving}. \\ 
\bottomrule
\end{tabularx}
\end{table*}

Table~\ref{tab:interaction-types} provides an overview of the 11 proposed types of developer-AI interactions, characterized by their triggers, the AI response, the developer reaction, the type of output generated, and concrete examples. The \textit{trigger} column describes how each interaction is initiated, ranging from automatic mechanisms such as typing context to explicit commands or workflow events. The \textit{AI response} column details the nature of the system's output, such as real-time suggestions, context-specific recommendations, or step-by-step explanations. The \textit{developer response} column captures how developers interact with and react to AI output, including actions such as reviewing, refining, or integrating suggestions. The \textit{output} column highlights the type of deliverables produced by AI, such as code snippets, documentation, or quality reports, offering insight into the specific contributions these interactions make to the development process. The \textit{example} column illustrates each interaction type with real-world applications, such as GitHub Copilot and ChatGPT. In the following, we briefly introduce each interaction type.

\paragraph{Auto-Complete Code Suggestions}
Auto-complete code suggestions in tools such as GitHub Copilot represent one of the most intuitive ways developers interact with AI, as these suggestions seamlessly integrate into the development workflow. Triggered automatically based on typing context, AI offers real-time recommendations in the form of ghost text or pop-ups, reducing the cognitive effort required for repetitive tasks. Developers can easily accept, scroll through, or dismiss these suggestions, allowing them to maintain focus while improving productivity.

\paragraph{Command-Driven Actions}

Command-driven actions allow developers to perform targeted tasks by issuing explicit instructions to the AI. These commands, such as generating documentation or summarizing code changes, empower developers to quickly access specific functionality. The AI processes these directives to produce customized outputs that the developers can review, edit, and refine. 

\paragraph{Conversational Assistance}

Conversational assistance enables developers to engage in natural language interactions with AI, making it an approachable collaborator. Developers can ask questions or request guidance on specific challenges, and the AI responds with explanations, suggestions, or even detailed code snippets. This interactive exchange facilitates learning, problem solving, and creativity. By mimicking human-like communication, this type of interaction reduces the cognitive burden of translating ideas into formal syntax, allowing developers to focus more on their core problem-solving objectives rather than wrestling with technical documentation or rigid query formats.

\paragraph{Contextual Recommendations}

Contextual recommendations leverage the AI’s ability to interpret cues from the project environment, providing developers with tailored suggestions that align with the specific context of their work. For example, the AI might suggest libraries, patterns, or improvements based on the file type, project dependencies, or overall structure. Unlike auto-complete code suggestions, which are triggered in real-time as the developer types and are focused on immediate, localized code completion, contextual recommendations take a broader view, offering guidance that aligns with the entire project or file. By focusing on project-wide context rather than line-by-line interactions, contextual recommendations complement autocomplete suggestions to provide more comprehensive support. Developers evaluate and integrate these suggestions as needed, ensuring that they align with the specific requirements of the task at hand.

\paragraph{Selection-Based Enhancements}

Selection-based enhancements focus on specific portions of code, allowing developers to receive customized assistance for highlighted segments. The AI responds by providing refactored code, generating test cases, or offering explanations for complex logic. Developers can then incorporate or adjust these enhancements, improving code quality and clarity. By narrowing its focus to specific selections, the AI ensures that its responses are relevant and actionable, streamlining tasks such as debugging and optimization.

\paragraph{Explicit UI Actions}

Explicit UI actions involve a more deliberate interaction, where developers manually trigger AI functionality through the tool interface. This might include running security scans, generating reports, or drafting documentation. The AI produces actionable outputs that developers can review, refine, and integrate into their workflow. This type of interaction emphasizes the importance of developer control, ensuring that AI support aligns precisely with their intentions.

\paragraph{Comment-Guided Prompts}

Comment-guided prompts allow developers to influence AI behavior using natural language descriptions embedded directly in the code as comments. The AI interprets these prompts to generate code snippets that align with the described intent. Developers can then review, adjust, and verify the generated output to ensure that it meets their requirements. 

\paragraph{Event-Based Triggers}

Event-based triggers enable automation during key workflow milestones, such as commits or pull requests. The AI responds to these events by performing predefined tasks such as running quality checks or security scans and producing actionable outputs such as reports or flagged issues. Developers then review these outputs to address any identified concerns. This interaction type is used, for example, to integrate AI into continuous integration and delivery pipelines, automating repetitive tasks and ensuring code quality.

\paragraph{Shortcut-Activated Commands}

Shortcut-activated commands streamline the interaction process by allowing developers to invoke AI functionality with predefined keyboard shortcuts. The AI responds with contextually relevant suggestions or documentation that developers can evaluate and incorporate into their work. This approach reduces disruption, allowing developers to maintain their workflow while accessing AI support quickly and efficiently. In contrast, explicit UI actions require more deliberate engagement through IDE buttons or menus.

\paragraph{File-Aware Suggestions}

File-aware suggestions provide developers with context-specific recommendations tailored to the type of file or directory they are working on. For example, the AI might suggest configuration options for a settings file or test templates for a testing script. Developers can review these suggestions and adapt them to fit their project requirements, ensuring that AI assistance is relevant and actionable. Note that file-aware suggestions provide recommendations tailored to specific file types at the time of file creation or editing. In contrast, contextual recommendations focus on broader project-level cues and suggest improvements during ongoing development.

\paragraph{Automated API Responses}

Automated API responses integrate AI capabilities into broader project workflows through API calls or webhooks. These interactions typically involve the generation of reports, release notes, or analysis outputs in response to predefined triggers. Developers review, modify, and integrate these outputs as needed, making this interaction type particularly useful for managing large-scale or repetitive tasks within collaborative projects. Automated API responses focus on system-to-system interactions triggered by webhooks or API calls, whereas event-based triggers are specific to workflow events like commits or pull requests within the developer's environment.

\section{Research Agenda}

The taxonomy presented in Section II provides a structured framework for understanding how developers interact with AI tools, serving as a foundation for identifying key research opportunities. Each type of interaction presents distinct challenges and areas for improvement. For instance, passive interactions such as code suggestions (e.g., GitHub Copilot) highlight questions about reducing cognitive load and balancing automation with developer control. Active interactions, such as conversational assistance or selection-based enhancements, raise additional considerations about trust, usability, and adaptability. These challenges and considerations form the basis for the research questions outlined in this agenda, which aim to align theoretical insights with practical advancements in developer-AI workflows.

Building on the taxonomy of developer-AI interaction types, this research agenda explores opportunities to optimize, expand, and tailor these interactions to better support developers. It highlights critical areas of study aimed at improving productivity, trust, customization, and usability in AI-driven software development tools.

\paragraph{Effectiveness of Interaction Types}

Different interaction types, such as auto-triggered suggestions, command-driven actions, and selection-based enhancements, have distinct impacts on productivity and code quality. Some types may better suit novice developers, while others might benefit experienced professionals. To explore these differences, we need to understand: Which interaction types lead to the most productive workflows in specific software development tasks? Are some types more suited to particular activities or user profiles? Comparative user studies that measure task completion times, accuracy, and satisfaction can help answer these questions. In addition, quantitative insights and qualitative feedback will reveal which types feel intuitive and effective in various scenarios.

\paragraph{Developer Trust and Adoption}

Trust is a crucial factor for developers when adopting AI tools, influenced by the transparency, reliability, and level of control they offer. Developers may approach high-stakes tasks, such as security scans, with skepticism, often requiring clear and reliable outputs to build confidence in the tool's capabilities. One question to investigate is how different types of interaction shape trust in AI outputs. Another is how interaction designs can foster trust without encouraging over-reliance or distrust. Surveys and interviews can uncover trust perceptions, especially in scenarios where AI modifies code. Behavioral experiments could track how often developers accept or override AI suggestions, offering insight into trust-related behaviors.

\paragraph{Context-aware AI Interactions}

AI tools often lack the ability to fully align their recommendations with the specific structure, goals, or context of a project. Incorporating project-level data, understanding dependencies, and recognizing design patterns could make AI suggestions more relevant and actionable. How could AI better adapt to the current task or the broader context of the project? Could integrating past interactions and project structure improve the relevance of suggestions? Developing context-aware models that embed these elements and comparing their acceptance rates with generic suggestions could provide valuable information.

\paragraph{Optimizing for Developer Control and Customization}

Striking the right balance between automation and developer control is vital. Features that allow developers to adjust the behavior of AI tools, such as the frequency or intrusiveness of suggestions, can help ensure that the tools remain helpful rather than overwhelming. Research in this area should consider how interaction designs empower developers without introducing unnecessary complexity. Questions such as ``What levels of customization do developers find most useful?'' and ``How can these features be implemented to enhance workflows?'' can guide this work. Testing customizable interfaces and surveying developers about their preferences can help refine these designs.

\paragraph{Reducing Cognitive Load}

AI tools can disrupt focus if their interactions are too frequent or intrusive. Minimizing cognitive load is essential for sustained productivity and engagement. A critical question is: What type of interaction best balances productivity and reduced cognitive burden? Research could compare passive or reactive interaction types to more intrusive ones. Methods such as eye tracking, task switching analysis, and subjective mental load assessments could help identify the least disruptive and most effective approaches.

\paragraph{Ethics and Bias in AI Interactions}

Ethical concerns and biases in AI interactions must be addressed to ensure fair and inclusive development practices. AI tools risk reinforcing stereotypes or perpetuating biases present in their training data. Investigating how AI suggestions differ according to developer characteristics or codebase patterns is an important step. What mechanisms can mitigate biases in AI outputs? Analyzing patterns across diverse developers and codebases can uncover problematic trends, while feedback systems that allow developers to flag issues can help refine AI behavior over time.

\paragraph{Privacy and Data Protection}

One critical area for future research is understanding and mitigating privacy and data protection concerns in developer-AI interactions. AI-powered tools often rely on sending contextual data, such as code snippets or project metadata, to external servers for processing. This raises questions about how much of a developer’s context is shared, who has access to the data, and how securely it is stored. Investigating privacy-preserving approaches, such as federated learning or on-device processing, could reduce the risks associated with exposing sensitive data. Future studies should evaluate trade-offs between the effectiveness of AI assistance and the privacy of developer workflows, exploring how these tools can operate without compromising security or confidentiality.

\paragraph{Hallucination and Damage Control}

Hallucinations in AI, where the system generates incorrect or misleading outputs, pose a significant challenge in developer workflows, especially when such outputs are integrated into critical systems. Research is needed to identify the causes of these hallucinations and to develop mechanisms for detecting and mitigating them. For example, integrating AI systems with feedback loops that allow developers to report inaccurate suggestions could improve the reliability of these tools over time. Furthermore, establishing clear safeguards, such as confidence scores or requiring explicit developer review before accepting high-risk changes, can minimize potential damage caused by erroneous AI outputs. 

\section{Conclusion}

This paper presents a taxonomy of the types of developer-AI interaction, providing a structured framework for understanding the diverse ways developers engage with AI-powered tools in software engineering. The accompanying research agenda builds on this taxonomy, identifying critical areas such as trust, cognitive load, and customization that warrant further exploration. Our work provides a foundation for understanding the diverse ways developers interact with AI-powered tools in software engineering. The taxonomy outlines key interaction types and dimensions, which can help guide further research and tool development, e.g., to better orchestrate various interaction types throughout the development workflow. Although validation is beyond the scope of this paper, future studies could empirically evaluate the taxonomy by observing tool use in real-world software development workflows. Future work could also include a systematic evaluation of existing tools to analyze the frequency and usage of each interaction type. In addition, researchers could explore how these interaction types align with developer needs in specific contexts, refining the taxonomy based on empirical findings.


\end{sloppy}
\end{document}